\documentclass[lettersize,journal]{IEEEtran}
\usepackage[utf8]{inputenc}
\usepackage[cmex10]{amsmath}
\usepackage{upgreek}
\usepackage{booktabs}
\usepackage{epsfig}
\usepackage{latexsym}
\usepackage{multirow}
\usepackage{stfloats}
\usepackage{epstopdf}
\usepackage{color}  
\usepackage{tabularx} 
\usepackage{algorithm}
\usepackage{algpseudocode}
\usepackage{amssymb}
\usepackage{enumerate}
\usepackage{array}
\graphicspath{{./Figures/}}
\usepackage{color}
\usepackage{bbm}
\usepackage{bm}
\usepackage{cite}
\usepackage[tight,footnotesize]{subfigure}
\usepackage{balance}
\usepackage{mathrsfs}
\usepackage{verbatim}
\usepackage{dsfont}
\usepackage{verbatim}
\usepackage{setspace}
\usepackage{diagbox}
\usepackage{multicol}
\usepackage{environ}
\usepackage{tikz}
\usepackage{amsmath}
\usepackage{stfloats}
\usepackage{algorithm}
\usepackage{algpseudocode}
\usepackage{amsmath}
\usepackage{graphics}
\usepackage{epsfig}
\usepackage{caption}
\allowdisplaybreaks[4]
\usepackage{amsmath}
\usepackage{amsthm}
\usepackage{authblk}

\def\BibTeX{{\rm B\kern-.05em{\sc i\kern-.025em b}\kern-.08em
		T\kern-.1667em\lower.7ex\hbox{E}\kern-.125emX}}
\begin{document}
\title{Joint Communication and Radar Sensing for Terahertz Space-Air-Ground Integrated Networks (SAGIN)
}
	
\author{Chong Han, Weijun Gao, Zhepu Yin, Chuang Yang, Mugen Peng, and Wenjun~Zhang



}
	\maketitle
	\thispagestyle{empty}
	\begin{abstract}
The transition from isolated systems to integrated solutions has driven the development of space-air-ground integrated networks (SAGIN) as well as the integration of communication and radar sensing functionalities. By leveraging the unique properties of the Terahertz (THz) band, THz joint communication and radar sensing (JCRS) supports high-speed communication and precise sensing, addressing the growing demands of SAGIN for connectivity and environmental awareness. However, most existing THz studies focus on terrestrial and static scenarios, with limited consideration for the dynamic and non-terrestrial environments of SAGIN.
In this paper, the THz JCRS techniques for SAGIN are comprehensively investigated. Specifically, propagation characteristics and channel models of THz waves in non-terrestrial environments are analyzed. A link capacity comparison with millimeter-wave, THz, and free-space optical frequency bands is conducted to highlight the advantages of THz frequencies. Moreover, novel JCRS waveform design strategies are presented to achieve mutual merit of communication and radar sensing, while networking strategies are developed to overcome challenges in SAGIN such as high mobility. Furthermore, advancements in THz device technologies, including antennas and amplifiers, are reviewed to assess their roles in enabling practical JCRS implementations.
	\end{abstract}
	
    \section{Introduction}
   
The growing demand for seamless connectivity and precise environmental awareness has driven the transition from isolated systems to integrated solutions. Traditionally, non-terrestrial platforms, e.g., satellites and unmanned aerial vehicles (UAVs), were developed for specific missions. However, with increasing complexity and scale, the space-air-ground integrated network (SAGIN) has emerged as a holistic approach for future networks. SAGIN facilitates collaboration among space-borne, airborne, and terrestrial nodes, enabling high-speed data exchange, global coverage, and improved operational efficiency.
Optical communication systems, as a conventional solution, can achieve ultra-high data rates but are highly susceptible to irradiance interference and Mie scattering or fog penetration, leading to significant performance degradation from sunlight and fog obstruction. To overcome this limitation, Terahertz (THz) band communications, operating in the 0.1–10 THz range, have been proposed as a promising alternative for fast and stable SAGIN wireless communications~\cite{akyildiz2022terahertz,nie2021channel,mao2022terahertz}.

Beyond high-speed data transmission, the THz band exhibits unique properties that make it highly suitable for radar sensing. 
First, over half of the total luminosity and the majority of photons emitted in the universe since the Big Bang fall within the THz and far-infrared regions, solidifying the importance of THz radar sensing for astronomical applications. Second, THz frequencies serve as a critical window for deep-space exploration, capturing thermal emissions from gas and dust at extremely low temperatures. Such emissions, typically peaking in the THz range, allow researchers to uncover hidden cosmic phenomena, like star formation and interstellar processes. Finally, THz radar systems are able to detect water vapor density profiles with high accuracy, facilitating atmospheric monitoring and weather prediction.
Therefore, although the idea of integrating communication and sensing is first proposed for 6G networks in terrestrial scenarios, THz joint communication and radar sensing (JCRS) is unprecedentedly interesting in THz SAGIN, which are vital pillars of next-generation networks\cite{elbir2024terahertz,alqaraghuli2023road,wu2022sensing,wu2023dft}. 


In this paper, we comprehensively investigate the enabling technologies and future prospects of THz JCRS techniques for SAGIN. First, we analyze the propagation characteristics of THz wave in various SAGIN scenarios, by considering key attenuation factors of THz signals, such as molecular absorption and atmospheric scattering. 
A comparative analysis of link capacity across millimeter-wave (mmWave/mmW), THz, and free-space optical (FSO) bands is presented, demonstrating the unique advantages of the THz spectrum for SAGIN applications.
Second, we explore novel waveform designs for THz JCRS systems, aiming to balance the dual requirements of high-speed communication and precise sensing, thereby maximizing the overall system efficiency. This includes discussions on waveform optimization strategies that address the trade-offs between bandwidth utilization and sensing accuracy. Furthermore, We examine recent advancements in THz hardware technologies, including high-gain antennas, efficient power amplifiers, and integrated circuits, highlighting their pivotal role in realizing practical THz-based SAGIN systems. Finally, we identify potential future research directions for THz JCRS in SAGIN to motivate future research studies in the field.

\section{Wave Propagation Analysis for Terahertz Integrated Communication and Radar Sensing}~\label{sec:channel}
\begin{figure*}
    \centering
    \includegraphics[width=\linewidth]{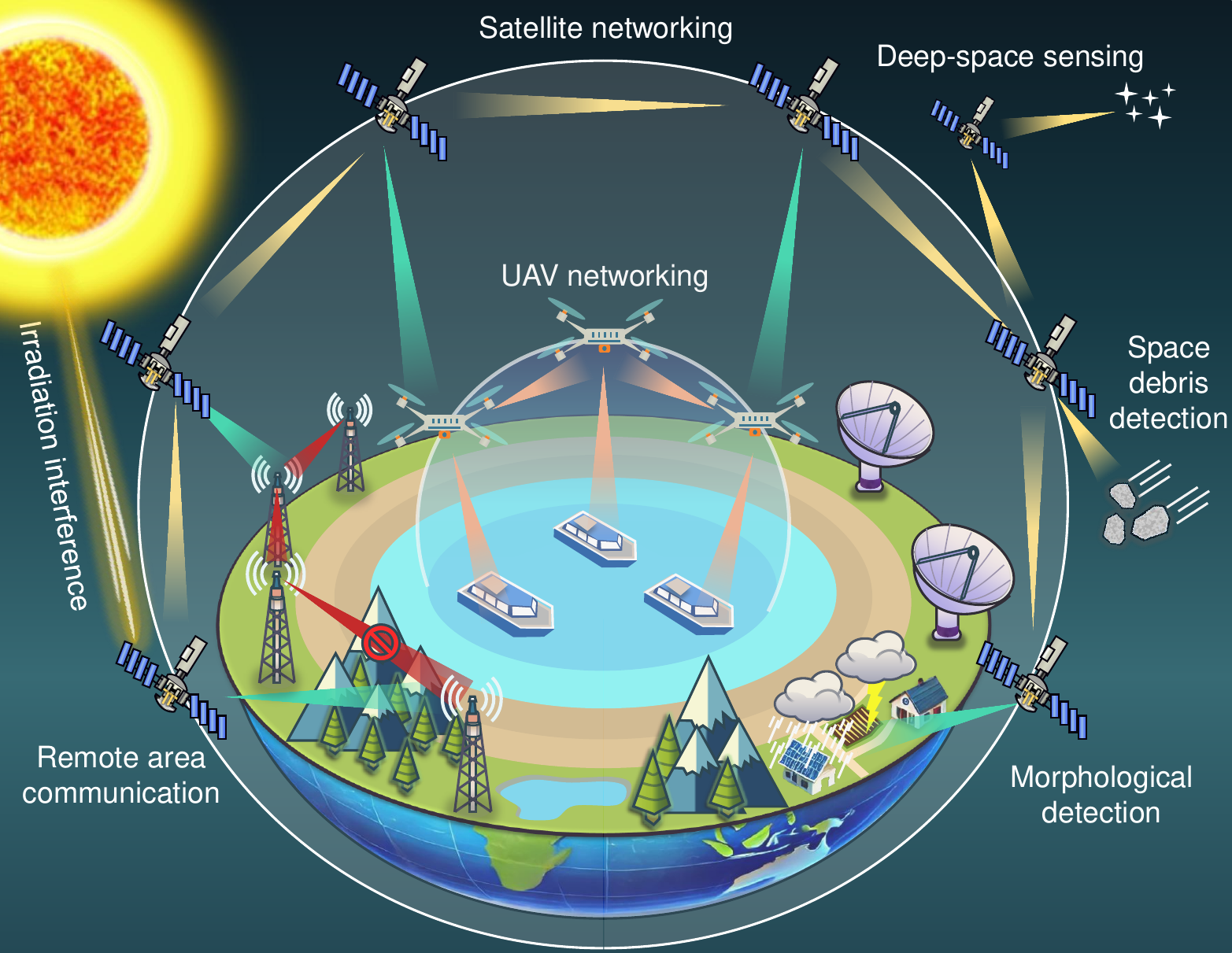}
    \caption{Illustration of THz SAGIN with JCRS applications.}
    \label{fig:system_model}
\end{figure*}
A representative vision of a future SAGIN enabled by THz wireless communications is depicted in Fig.~\ref{fig:system_model}. THz-based SAGIN encompasses diverse wave propagation scenarios, such as air-to-ground, air-to-air, satellite-to-ground, satellite-to-air, and inter-satellite links. The propagation characteristics vary significantly across altitudes due to distinct medium components. This complexity makes wave propagation analysis for THz SAGIN more challenging than for terrestrial systems. Understanding how THz electromagnetic (EM) waves interact with the external environment is essential to address these challenges.
\subsection{Terahertz Wave Propagation Modeling in SAGIN}
Unlike terrestrial scenarios where lots of obstacles and reflectors create multi-paths for wave propagation, the number of multi-paths in non-terrestrial THz wave propagation is small. Therefore, we usually only need to consider one line-of-sight path from the transmitter to the receiver. This is verified in a study on UAV multi-path analysis~\cite{li2021ray} that the K-factor, characterizing the power ratio of LoS path to other multi-paths, is greater than 50 at an altitude higher than $50~\textrm{m}$. 

A universal THz wave attenuation model for SAGIN has been proposed in~\cite{yang2024universal}, where the attenuation is modeled as the result of a collision between THz photons and medium particles, and the absorption and scattering effects led by various medium particles are elaborated. Based on this model, the modeling of THz wave propagation in SAGIN can be divided into two sub-problems, i.e., modeling the cross-section of various medium particles in the THz band characterizing the intensity of absorption and scattering, and modeling their corresponding number density along the propagation path.  
The specific THz wave attenuation model is described as follows:
\subsubsection{Molecular absorption effect} 
Air molecule components including oxygen, nitrogen, and water vapor lead to the absorption of EM save, which is called the molecular absorption effect. As comprehensively modeled in~\cite{jornet2011channel}, the molecular absorption effect is the result of the energy level transition of molecules after the collision between THz photons and molecules, according to quantum mechanisms. Only molecules with equal intrinsic energy levels different from THz photon energy can absorb the photon, leading to a large absorption cross-section. By jointly considering the absorption cross-section and the number density in the atmosphere, it is revealed that water vapor has a significant absorption effect with about six orders of magnitude larger than other air components. 
Altitude is another important factor affecting molecular absorption loss. As altitude increases, the density of water vapor decreases, which shows an exponential relationship with altitude. This implies that the molecular absorption effect becomes much weaker at high altitudes. It is revealed that at an altitude higher than about $10~\textrm{km}$, the molecular absorption effect becomes very weak due to the low number density of water vapor. Therefore, this property can be leveraged in SAGIN to reduce interference between space-borne and terrestrial nodes.
\subsubsection{Weather effects}
\begin{figure}
    \centering
    \includegraphics[width=\linewidth]{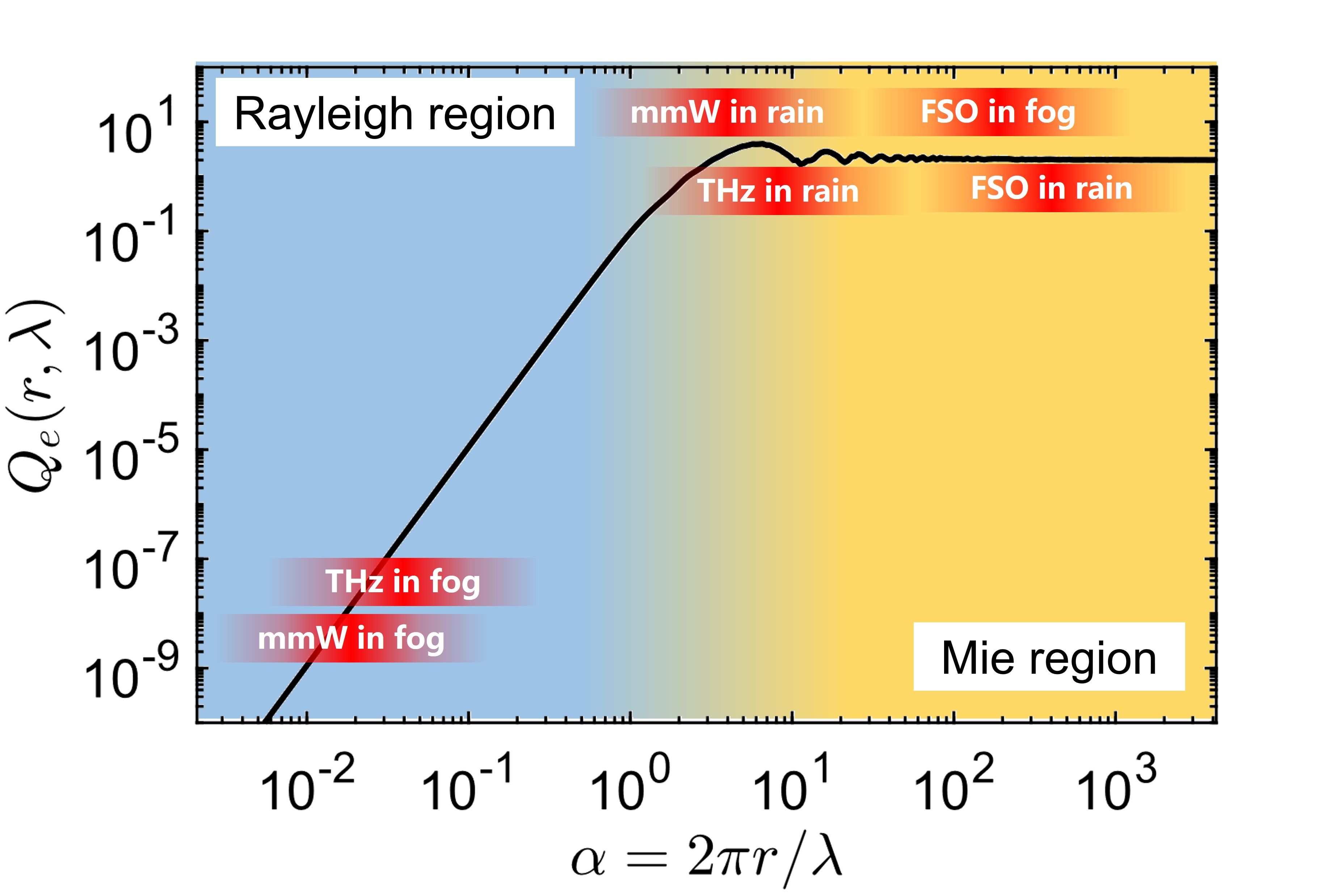}
    \caption{Extinction cross section of medium particles in rain and fog for different EM waves in the mmWave, THz, and FSO frequency bands. The typical radius of rain droplets and fog droplets is $2~\textrm{mm}$ and $20~\mu\textrm{m}$, respectively. Typical frequencies/wavelengths for the mmWave, THz, and FSO frequency bands are 0.02 THz, 0.3 THz, and 1550 nm, respectively.
    \textit{Rayleigh region} (highlighted in blue) corresponds to conditions that $r\ll \lambda$, specifically $\alpha< 0.1$. \textit{Mie region} (highlighted in yellow) occurs when the particle size is comparable to or larger than the wavelength, i.e., $\alpha\ge 0.1$, where $Q_e(r,\lambda)$ is on the order of $10^0~\textrm{m}^2$.
    }
    \label{fig:mie-comparison}
\end{figure}

Due to the atmospheric circulation, various weather conditions including rain, dust, fog, and snow appear in the troposphere, leading to different attenuation effects. Unlike molecules in the atmosphere whose cross sections depend on energy level difference, these large-scale medium particles heavily rely on their particle radius, which can be either smaller or on the same order of magnitude as THz wavelengths. For example, radii of raindrops are about $1\textrm{mm}$ to $10~\textrm{mm}$, which is on the same order of magnitude of $0.3~\textrm{THz}$ EM wave, while radii of fog droplets are around several micrometers, which is much smaller than THz wavelengths.
Mie absorption and scattering models are applied to characterize the cross sections when the particle radius is on the same order of magnitude to THz wavelengths, or otherwise, Rayleigh ones are used. As demonstrated in Fig.~\ref{fig:mie-comparison}, THz wave propagation can be severely scattered by rain droplets but not significantly affected by fog. 

\begin{figure*}
    \centering
    \includegraphics[width=0.8\linewidth]{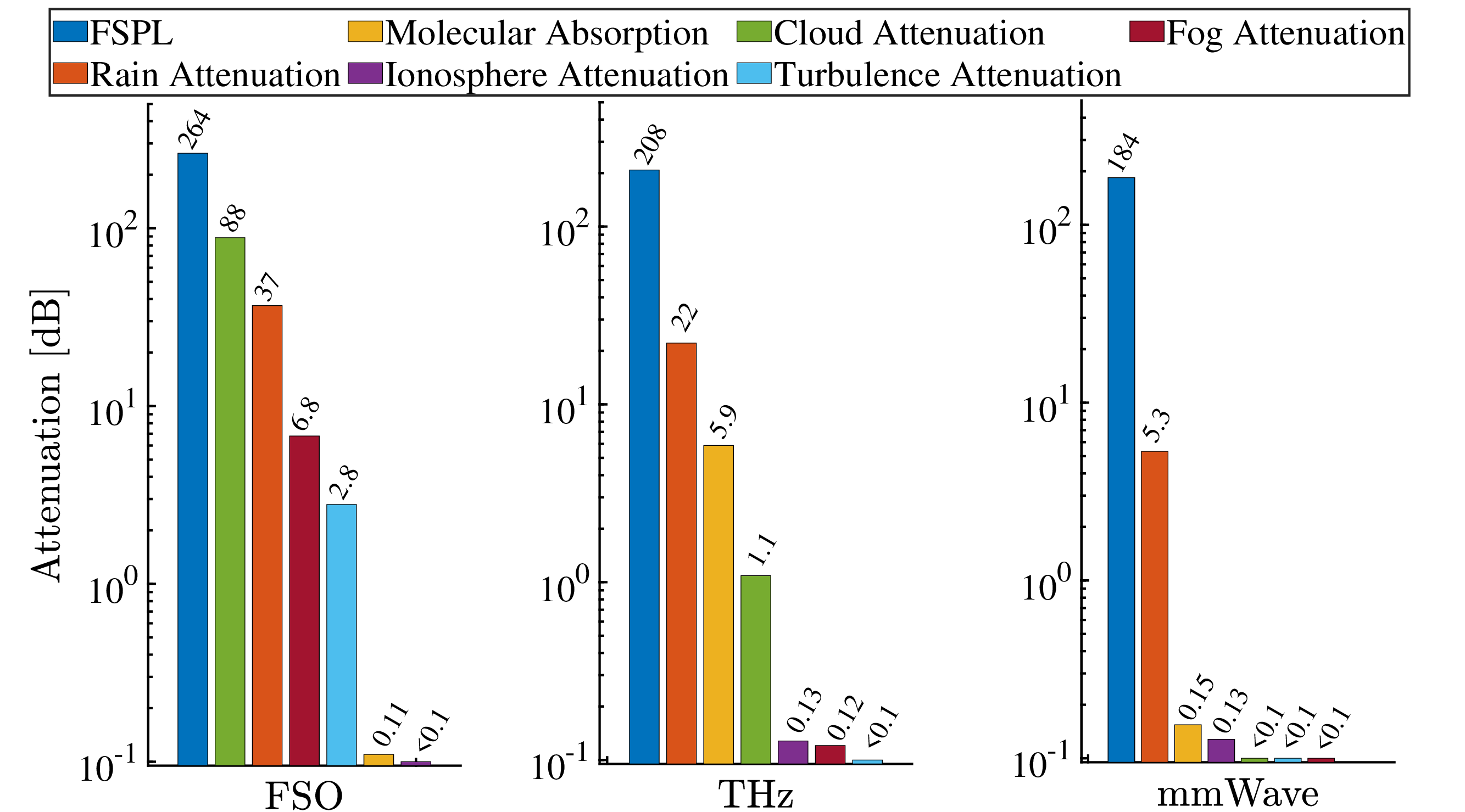}
    \caption{Comparison of different attenuation factors in a 2000-km ground-to-satellite link for the FSO, THz, and mmWave frequency bands, in terms of FSPL, molecular absorption, cloud, fog, rain, ionosphere, turbulence, and Mie scattering. Attenuation factors with less than $0.1~\textrm{dB}$ loss are marked with $<0.1$.} 
    \label{fig:S2A_component}
\end{figure*}
\begin{figure*}
    \centering
    \includegraphics[width=0.8\linewidth]{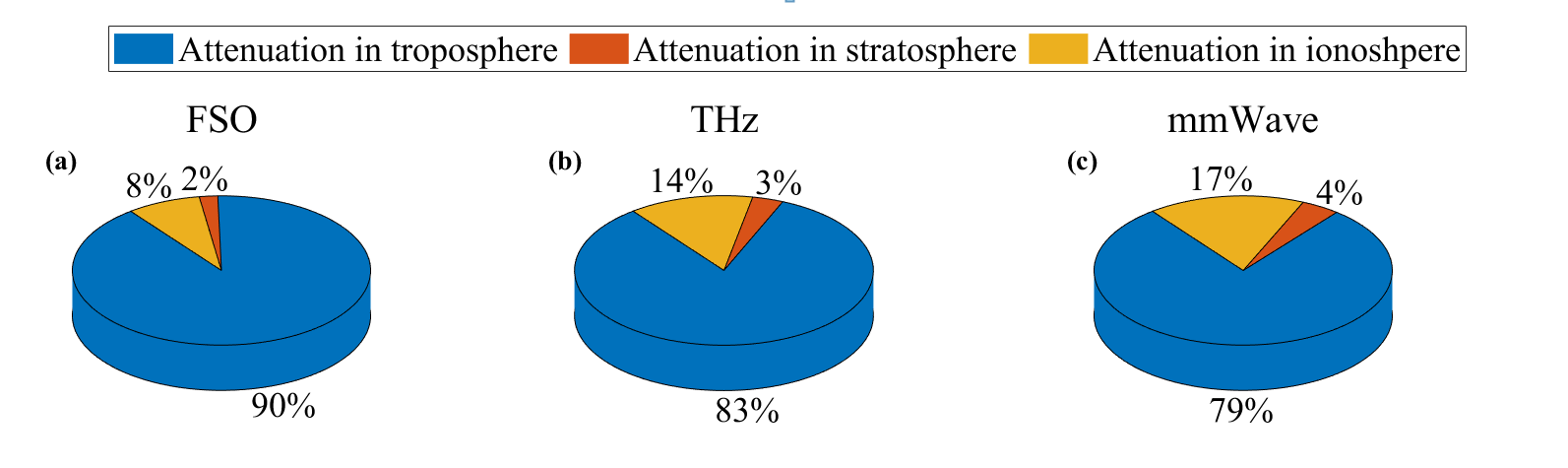}
    \caption{Comparison of attenuation contributions from the troposphere, stratosphere, and ionosphere in a 2000-km ground-to-satellite link for the FSO, THz, and mmWave frequency bands.} 
    \label{fig:percentage}
\end{figure*}

\subsubsection{Turbulence attenuation}
Random airflow causes chaotic atmospheric turbulence, which is a major attenuation factor for FSO communications. However, as analyzed in~\cite{gao2024attenuation}, lower frequencies experience significantly less turbulence attenuation (proportional to $f^{7/6}$). Consequently, the influence of atmospheric turbulence on THz wave propagation can generally be neglected, except under extreme conditions.

\subsubsection{Plasma attenuation}
The attenuation of plasma in the Ionosphere results from three phenomena~\cite{nie2021channel}, i.e., collision between THz waves and particles, oscillation resonance of electric field of free electrons, and Faraday rotation leading to depolarization of THz wave. However, it is demonstrated that the attenuation factor in the Ionosphere is less than $10^{-3}~\textrm{dB}/\textrm{km}$ and therefore can be omitted.

\subsection{Capacity Comparison for Millimeter-Wave, Terahertz, and Free-space Optical Band}
\begin{figure*}[t]
    \centering
    \includegraphics[width=0.8\linewidth]{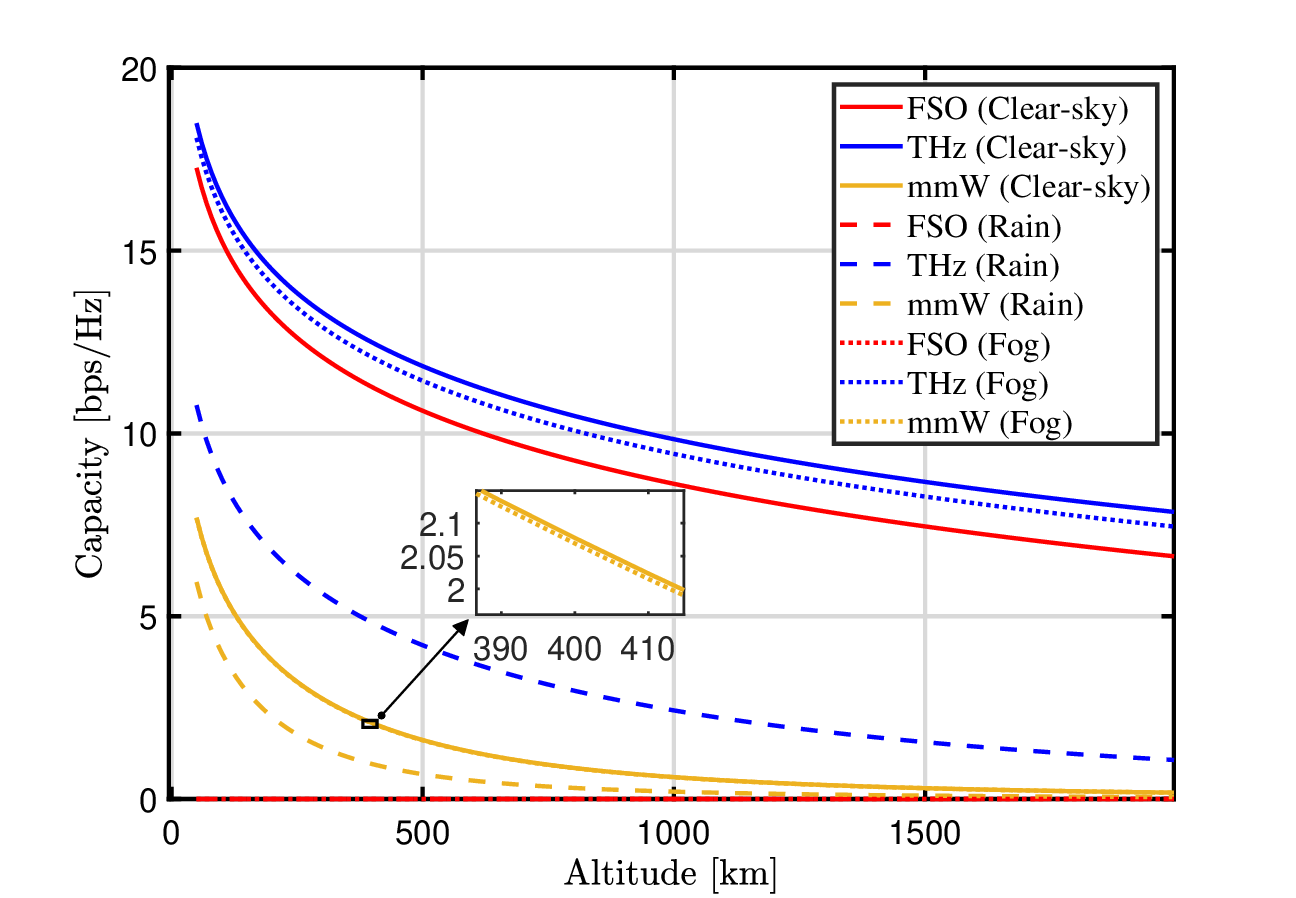}
    \caption{Comparison of communication capacity of a satellite-to-ground wireless link at different frequency bands in various weather conditions. The specific parameters for the three frequency bands are summarized in Table.~\ref{tab:link_budget_params}.}
    \label{fig:S2A_capacity}
\end{figure*}

To demonstrate the significance of applying THz wireless communications against conventional ones, we compare the capacity of a ground-to-satellite link for the mmWave, THz, and FSO bands. 
As summarized in Fig.~\ref{fig:S2A_component}, the attenuation factors for a ground-to-satellite link across FSO, THz, and mmWave frequency bands are compared. We observe that for all three frequency bands, the free-space path loss (FSPL) is the most dominant attenuation factor among all factors, and as frequency increases, the FSPL becomes larger. Table.~\ref{tab:link_budget_params} describes the parameters for the link budget analysis among three frequency bands, based on which the communication capacity is calculated and plotted in Fig.~\ref{fig:S2A_capacity}.
Detailed observations and comparisons among these frequency bands are summarized as follows:
\begin{itemize}
    \item \textbf{FSO}: FSO communication exhibits the highest free-space path loss (FSPL) among the three frequency bands, reaching approximately $264~\textrm{dB}$. In addition, atmospheric factors such as cloud, rain, and fog impose severe attenuation, with cloud attenuation being the second largest contributor at $88~\textrm{dB}$. Despite these challenges, FSO links benefit from extremely high transmit and receive  antenna gains (exceeding $100~\textrm{dBi}$), making them suitable for high-altitude applications, particularly in clear sky conditions. 
    \item \textbf{THz}: he THz band exhibits a lower FSPL of around $208~\textrm{dB}$ compared to FSO. Apart from FSPL, rain attenuation and molecular absorption dominate the overall path loss, contributing $22~\textrm{dB}$ and $5.9~\textrm{dB}$, respectively. Notably, molecular absorption loss remains moderate because water vapor density decreases exponentially with altitude. In the stratosphere and beyond, molecular absorption becomes negligible. Furthermore, despite the propagation losses, the THz band provides the highest link capacity (bps/Hz) under various atmospheric conditions, including clear sky, rain, and fog.  
    \item \textbf{mmWave}:  The mmWave band experiences the lowest FSPL among the three, measuring approximately $184~\textrm{dB}$. Additionally, its atmospheric attenuation is significantly lower compared to FSO and THz. Rain attenuation remains the most significant factor at $5.3~\textrm{dB}$, while molecular absorption and other weather-related losses remain below $0.2~\textrm{dB}$. This makes mmWave a viable option for space-air-ground communications, offering a balance between manageable attenuation and reasonable link capacity. 
\end{itemize}
As shown in Fig.~\ref{fig:percentage}, the total attenuation in a satellite-to-ground link can be decomposed into contributions from different atmospheric layers, namely the troposphere, stratosphere, and ionosphere. The results indicate that attenuation in the troposphere is the most significant across all three communication technologies, including FSO, THz, and mmWave, accounting for more than 79\% of the total loss. This is primarily due to the presence of water vapor, clouds, and weather-induced variations in this layer, which have a strong impact on signal propagation, especially at higher frequencies. In contrast, attenuation in the stratosphere is relatively small, contributing only 2\% to 4\% of the total propagation loss since the stratosphere is almost dry and stable. However, attenuation in the ionosphere is higher than in the stratosphere, contributing 8\% to 17\% of the total loss, particularly for lower-frequency signals such as mmWave. This is mainly due to the longer propagation distance through the ionosphere and the influence of free electrons, which can cause dispersion and scintillation effects. The comparison across different frequency bands highlights the dominant role of tropospheric conditions in determining overall link performance. This underscores the importance of adaptive channel modeling and mitigation strategies in practical system designs.


\begin{table*}[t]
\centering
\caption{Parameters for the ground-to-satellite link for mmW, THz, and FSO frequency bands.} 
\includegraphics[width=0.8\textwidth]{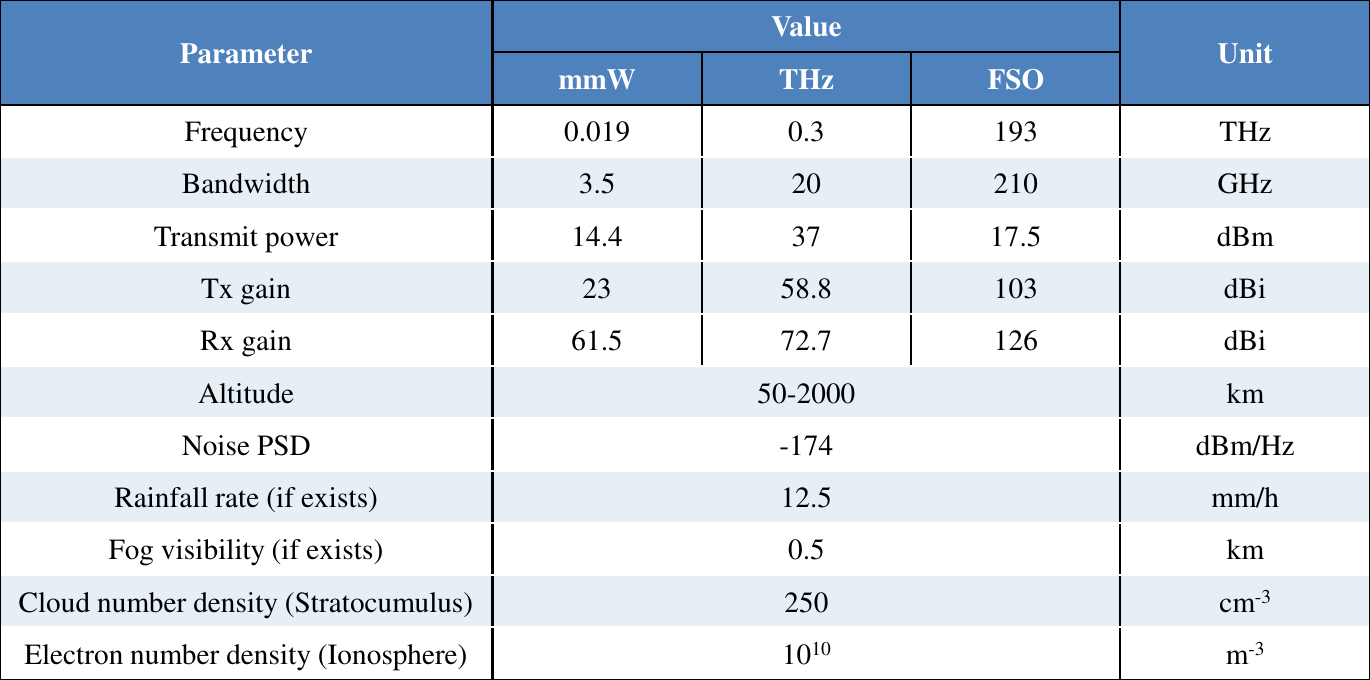} 
\label{tab:link_budget_params} 
\end{table*}

\section{Waveform Design for Terahertz Integrated Communication and Radar Sensing}~\label{sec:JCRS}
One of the main motivations for applying THz wireless communications to SAGIN is to establish backhaul links for sensing satellites while maintaining radar sensing capabilities. Integrating communication and radar sensing functionalities enables device miniaturization and scaling, which are crucial for modern space-air-ground applications. 
However, communication and sensing waveforms have traditionally been designed for separate objectives: communication waveforms aim to maximize data throughput while sensing waveforms focus on accurate target detection. Achieving JCRS requires innovative waveform designs capable of balancing these goals effectively.

\subsection{Communication-centric Waveform}
Orthogonal frequency division multiplexing (OFDM) is a widely used waveform in communication systems, particularly in 4G networks. By multiplexing multiple sub-carriers, OFDM achieves high spectral efficiency and reduces inter-carrier interference (ICI) under stationary conditions. However, in THz SAGIN, conventional OFDM faces significant challenges:
\begin{itemize}
    \item \textbf{High Peak-to-Average Power Ratio (PAPR):}
The high PAPR of OFDM leads to inefficient power utilization, particularly in THz systems where power amplifiers are constrained by hardware limitations in satellites and UAVs. This reduces the average capacity and overall system efficiency.
    \item \textbf{Strong Doppler Effect:} 
The high mobility of nodes, such as satellites and UAVs, introduces varying levels of Doppler effects. Satellites, moving at speeds up to 7.5 km/s, experience stronger Doppler shifts compared to UAVs, which typically move at speeds of 50–100 m/s. Additionally, fast-moving sensing targets like space debris further exacerbate Doppler-induced distortions, making OFDM unsuitable for these dynamic environments.
\end{itemize}
To address these issues, advanced OFDM-based waveforms have been proposed. Orthogonal time frequency space (OTFS) waveform transforms signals into the delay-Doppler domain, offering robustness against Doppler effects in high-mobility scenarios. Another improvement involves applying a Discrete Fourier Transform (DFT) spreading module before the inverse fast Fourier transform (IFFT). This spreads one symbol across all sub-carriers, significantly reducing the PAPR. By combining these techniques, the DFT-spread OTFS waveform achieves reduced PAPR and enhanced Doppler resilience, making it suitable for THz SAGIN. However, this comes at the cost of increased system complexity and reduced processing efficiency, particularly in real-time applications where high computational demands may limit practical performance.

\subsection{Sensing-centric Waveform}
Conventional sensing waveforms, such as frequency-modulated continuous waveform (FMCW), are widely used in radar systems due to their chirp-based nature. A chirp signal is characterized by a frequency that linearly increases or decreases over time, making it highly suitable for sensing as it allows precise estimation of range and velocity. However, FMCW is spectrally inefficient, as chirp signals occupy a large bandwidth that is not fully utilized for communication purposes, posing challenges for JCRS systems.

Affine Frequency-Division Multiplexing (AFDM) offers a sensing-centric solution by embedding communication data into the sensing waveform through delay-Doppler orthogonality. Utilizing the inverse discrete affine Fourier transform (IDAFT), AFDM efficiently adapts to both delay and Doppler spread conditions. This enables AFDM to optimize spectral resources without compromising sensing resolution, achieving precise range and velocity estimation while simultaneously supporting high-speed communication. These features make AFDM particularly suitable for JCRS applications, where simultaneous communication and sensing are essential.

Besides the aforementioned waveform technologies, THz wavefront engineering techniques, such as orbital angular momentum, Bessel beams, and Airy beams are alternative approaches for enhancing communication performance~\cite{gao2025channel}. These techniques manipulate the spatial properties of the THz waves to achieve better control of signal propagation. Moreover, combining waveform and wavefront design allows for a more integrated approach, where the waveform optimizes signal shaping for the channel, while wavefront engineering fine-tunes the spatial characteristics of the signal. This synergy improves both capacity and reliability, offering a promising solution for dynamic and high-mobility communication environments.

\section{Device Technology and Prototype}\label{sec:device}
The development of robust, efficient, and compact THz devices is crucial for realizing JCRS functionality in SAGIN, which faces several challenges, including parasitic effects at high frequencies, fabrication constraints, and radiation-induced degradation from gamma rays and proton irradiation. Furthermore, the robustness against extreme environmental conditions, such as severe temperature fluctuations and exposure to solar winds, is also essential for ensuring reliable operation in space. To overcome these challenges, advancements in semiconductor technologies and integrated antenna systems are vital for enabling high-speed, precise, and energy-efficient JCRS systems in diverse SAGIN scenarios.

\subsection{Advancements in Semiconductor Devices}
Conventional electronic components, such as transistors and diodes, face significant performance limitations at THz frequencies due to enhanced parasitic effects. Among these, parasitic capacitance is a critical issue, as it leads to signal distortion and power loss by storing and dissipating energy in unintended areas of the circuit. This problem is amplified by the shorter wavelengths and higher frequencies in THz systems, severely impacting device gain and frequency response.

To address these issues, III-V semiconductors, such as GaN and InP, have emerged as promising materials for THz devices. Their superior electron mobility and high breakdown voltage enable the development of high-efficiency THz power amplifiers and frequency multipliers. Additionally, III-V semiconductors offer inherent resistance against radiation-induced damage, making them ideal for spaceborne applications. For instance, GaAs-based devices demonstrate robust performance under proton irradiation, with minimal displacement damage, while effective shielding and advanced fabrication techniques mitigate gamma-ray-induced ionization effects. These advancements support the development of compact, high-performance THz devices capable of delivering multi-gigabit-per-second communication over multi-kilometer distances.

\subsection{Existing Prototype and System}
Recently proposed THz wireless communication systems mostly focus on terrestrial scenarios, aiming at achieving higher data rates or reaching longer communication distances. For instance, a 0.22 THz communication system with an antenna gain of 55 dBi with a 45 cm Cassegrain antenna is proposed, enabling an 84 Gbps data rate over 1.26 km~\cite{liu2024high}. However, such designs are based on large-size and static antenna systems, e.g., lens antenna, and are not suitable for high-mobility environments like SAGIN. 

To address the limitations, ongoing research is focusing on spaceborne THz communication applications, such as CubeSat platforms, which aim to achieve lightweight, compact, and efficient antenna designs while ensuring high performance in the challenging space environment~\cite{aliaga2022joint}. Current advancements emphasize integrating III-V-based amplifiers and frequency multipliers with reconfigurable antenna arrays to enhance the adaptability of the system, allowing for dynamic beam steering and tracking in spaceborne communication systems. 

\section{Future Directions}~\label{sec:future}
While significant progress has been made in the integration of THz communication and radar sensing within the SAGIN framework, several challenges and opportunities remain. The following future research directions are identified for THz JCRS in SAGIN.
\subsection{Performance Trade-off Between JCRS}
The performance trade-off, specifically between communication capacity and sensing Cramér-Rao Bound (CRB), remains a critical aspect of JCRS for THz SAGIN. Analyzing this trade-off is essential for optimizing joint system performance. Previous studies, such as~\cite{xiong2023fundamental}, have extensively explored this trade-off in Gaussian channels by examining the CRB-rate region. However, in the context of THz SAGIN, this trade-off is fundamentally different due to the sparsity inherent to THz propagation, which necessitates a re-evaluation of the CRB-rate region under THz-specific conditions, incorporating factors such as atmospheric attenuation, molecular absorption, and Doppler effects caused by the high mobility of satellites and UAVs.
Moreover, the integration norm of THz JCRS, including waveform design, antenna architecture, and system hardware optimization, should align with the unique THz channel characteristics. More approaches to joint waveform design that simultaneously optimize communication capacity and sensing accuracy are required. 
\subsection{AI-empowered Communication-Sensing-Computing Joint Network}
In addition to communication and sensing, computing plays a critical role in ensuring the efficient operation of SAGIN, given the computational constraints of satellites and UAVs. Jointly optimizing communication, sensing, and computing functions within a unified framework is one of the future directions.
On one hand, communication and sensing functionalities require significant computational resources for tasks such as signal processing, beamforming, and data fusion. Simultaneously, edge computing, as a distributed approach across SAGIN infrastructures, is essential for managing complex missions. This demands seamless integration with high-speed wireless communication.

AI-driven solutions offer transformative potential for addressing these needs. With the aid of AI-empowered networks, the joint network is able to dynamically optimize resource allocation including bandwidth, energy, and computational power. For example, machine learning models can predict network traffic and sensing demands to enable proactive resource management. Federated learning frameworks can train AI models collaboratively across distributed infrastructures, ensuring data privacy and reducing communication overhead. Moreover, advanced AI algorithms are essential for managing task scheduling, optimizing routing, and maintaining reliability in the dynamic and resource-constrained SAGIN environment.
\subsection{JCRS for Space Debris Detection}
Space debris detection is emerging as a potential application of JCRS in SAGIN. Space debris poses severe threats to satellites, spacecraft, and other critical assets in orbit. Conventional approaches to space debris detection, e.g., ground-based radar and optical systems, face limitations in detecting small-sized objects. THz JCRS, leveraging its short wavelengths, offers a promising solution. THz-based sensing provides centimeter-level resolution, enabling the detection and tracking of debris objects with a diameter less than $10~\textrm{cm}$. By integrating communication and radar sensing capabilities, JCRS can realize data transfer and space debris detection simultaneously. 
To achieve effective debris detection, tailored THz JCRS waveforms are required to balance communication and sensing functions while overcoming challenges including THz wave attenuation and Doppler effect due to the high mobility of debris. Moreover, AI-driven algorithms are envisioned to enhance debris identification by analyzing radar signatures and distinguishing between debris and other objects. The application of JCRS for space debris detection represents a critical step toward ensuring the safety and sustainability of space operations, and it highlights the broader potential of SAGIN-enabled technologies.
\subsection{MAC Protocol and Networking}\label{sec:MAC}
Link-layer MAC protocols and networking-layer designs are critical to addressing the challenges of the dynamic topology of THz SAGIN. Unlike conventional MAC protocol design for omnidirectional and low-frequency communications, THz communication faces challenges such as highly directional beams and sensitivity to environmental conditions, which complicate node discovery, beam alignment and tracking, and synchronization. On one hand, sensing-assisted MAC protocols, which leverage radar sensing data such as precise location and environmental feedback, hold promise for overcoming these communication challenges. On the other hand, radar sensing performed by satellites or space stations requires ultra-high-speed wireless links for sensing information backhaul. Future research direction includes designing THz protocols to make communication and radar sensing functionalities mutually aid each other. 

At the networking layer, innovative designs are required to manage the dynamic and heterogeneous topology of SAGIN. The rapid movement of satellites and UAVs, combined with the directional constraints of THz links, necessitates robust routing and resource allocation mechanisms. Future research could explore deep reinforcement learning (DRL)-based algorithms to dynamically optimize resource allocation, balancing throughput, latency, and energy efficiency. Additionally, adaptive routing protocols capable of maintaining link reliability under rapid topology changes are critical. By integrating advancements in MAC protocols and networking designs, future work will provide a unified framework to support THz JCRS in SAGIN, meeting the demands of next-generation communication and sensing applications.
\section{Conclusion}\label{sec:concl}
In this paper, we explore the integration of THz communication and radar sensing within the SAGIN framework, reflecting two major trends: The transition from isolated systems to integrated networks and the convergence of communication and sensing functionalities.  By leveraging the unique characteristics of the THz band in SAGIN, THz JCRS needs to address challenges in high-speed communication and precise sensing within dynamic environments. The presented methods offer tailored approaches for waveform design and device technology, as key enablers of THz JCRS techniques.
Future works focus on analyzing the trade-offs between communication capacity and sensing accuracy under THz-specific conditions, developing AI-driven communication-sensing-computing joint networks to meet the growing demands of SAGIN operations, and designing MAC and network layer protocols. This article highlights the transition from isolated systems to fully integrated solutions, paving the way for the next generation of intelligent and interconnected networks.
\bibliographystyle{IEEEtran}
\bibliography{ISAC}

\section{Biographies}

\vspace{-33pt}
\begin{IEEEbiographynophoto}{Chong~Han}
is with the Terahertz Wireless Communications (TWC) Laboratory, Department of Electronic Engineering and Cooperative Medianet Innovation Center (CMIC), Shanghai Jiao Tong University, Shanghai 200240, China (email:~chong.han@sjtu.edu.cn). 
\end{IEEEbiographynophoto}
\vspace{-33pt}
\begin{IEEEbiographynophoto}{Weijun~Gao}
is with the Terahertz Wireless Communications (TWC) Laboratory, Shanghai Jiao Tong University, Shanghai 200240, China (email:~gaoweijun@sjtu.edu.cn).
\end{IEEEbiographynophoto}
\vspace{-33pt}
\begin{IEEEbiographynophoto}{Zhepu~Yin}
is with the Terahertz Wireless Communications (TWC) Laboratory, Shanghai Jiao Tong University, Shanghai 200240, China (email:~zhepu.yin@sjtu.edu.cn).
\end{IEEEbiographynophoto}
\vspace{-33pt}
\begin{IEEEbiographynophoto}{Chuang~Yang}
is with the State Key Laboratory of Networking and Switching Technology, Beijing University of Posts and Telecommunications, Beijing 100876, China (e-mail:~chuangyang@bupt.edu.cn).
\end{IEEEbiographynophoto}
\vspace{-33pt}
\begin{IEEEbiographynophoto}{Mugen~Peng}
is with the State Key Laboratory of Networking and Switching Technology, Beijing University of Posts and Telecommunications, Beijing 100876, China (email:~pmg@bupt.edu.cn). 
\end{IEEEbiographynophoto}
\vspace{-33pt}
\begin{IEEEbiographynophoto}{Wenjun~Zhang}
is with Department of Electronic Engineering and Cooperative Medianet Innovation Center (CMIC), Shanghai Jiao Tong University, Shanghai 200240, China (email:~zhangwenjun@sjtu.edu.cn).
\end{IEEEbiographynophoto}
\vfill
\end{document}